\documentclass[fleqn,usenatbib]{mnras}

\usepackage[T1]{fontenc}

\DeclareRobustCommand{\VAN}[3]{#2}
\let\VANthebibliography\thebibliography
\def\thebibliography{\DeclareRobustCommand{\VAN}[3]{##3}\VANthebibliography}

\usepackage{graphicx}	
\usepackage{amsmath}	
\usepackage{amssymb}	
\usepackage{newtxtext,newtxmath} 

\newcommand{\approxcorr}{\mathop{\hat{\approx}}}

\newcommand{\dv}[2]{\frac{\mathrm{d} #1}{\mathrm{d} #2}}
\newcommand{\dvtw}[2]{\frac{\mathrm{d}^2 #1}{\mathrm{d} #2 ^2}} 
\newcommand{\fdv}[2]{\frac{\delta #1}{\delta #2}}
\newcommand{\pdv}[2]{\frac{\partial #1}{\partial #2}}
\newcommand{\D}{\text{d}}

\title[Cosmic Shear Sensitivity]{On the sensitivity of weak gravitational lensing to the cosmic expansion function} 

\author[Christian F. Schmidt, Matthias Bartelmann]{
Christian F. Schmidt,$^{1,2}$\thanks{E-mail: christian.schmidt@uni-jena.de}
Matthias Bartelmann,$^{2}$\thanks{E-mail: bartelmann@uni-heidelberg.de} \\
$^{1}$Institute for Theoretical Physics, Jena University, Froebelstieg 1, 07743 Jena, Germany \\
$^{2}$Institute for Theoretical Physics, Heidelberg University, Philosophenweg 16, 69120 Heidelberg, Germany}

\date{Accepted XXX. Received YYY; in original form ZZZ}

\pubyear{2023}

\begin{document}
\label{firstpage}
\pagerange{\pageref{firstpage}--\pageref{lastpage}}
\maketitle

\begin{abstract}
We analyse the functional derivative of the cosmic-shear power spectrum $C_\ell^\gamma$ with respect to the cosmic expansion function. Our interest in doing so is two-fold: (i) In view of attempts to detect minor changes of the cosmic expansion function which may be due to a possibly time-dependent dark-energy density, we wish to know how sensitive the weak-lensing power spectrum is to changes in the expansion function. (ii) In view of recent empirical determinations of the cosmic expansion function from distance measurements, independent of specific cosmological models, we wish to find out how uncertainties in the expansion function translate to uncertainties in the cosmic-shear power spectrum. We find the following answers: Relative changes of the expansion function are amplified by the cosmic-shear power spectrum by a factor {{$\approx 2-6$}}, weakly depending on the scale factor where the change is applied, and the current uncertainty of one example for an empirically determined expansion function translates to a relative uncertainty of the cosmic-shear power spectrum of {{$\approx10\,\%$}}.
\end{abstract}




\section{Introduction}

Mainly owing to its conceptual simplicity, gravitational lensing has developed into one of the most informative and reliable methods of observational cosmology \citep{2001PhR...340..291B, schneider2006gravitational, 2010CQGra..27w3001B, 2015RPPh...78h6901K, 2018ARA&A..56..393M}. Expected weak-lensing power spectra depend on the cosmological background model in two ways: geometrically via the angular-diameter distances entering its geometrical weight function, and dynamically via the growth of density perturbations. In view of these dependences, we address in this paper the following question:

If the cosmic expansion function $E(a) = H(a)/H_0$ is varied in an arbitrary way, how does the power spectrum of cosmological weak lensing change? Here, $H(a)$ is the Hubble function and $H_0$ its present value. We believe that this question is relevant for two main reasons: First, it is interesting to find out how sensitive the weak-lensing power spectrum $C_\ell^\gamma$ is to changes of the expansion function as a function of redshift. Or, phrased differently, at what redshift is $C_\ell^\gamma$ most or least sensitive to uncertainties in or modifications of $E(a)$? This question is particularly important in view of a possible time dependence of the dark energy \citep{2018LRR....21....2A, 2018RPPh...81a6902B}. Second, model-independent or, perhaps more appropriately, empirical methods of constraining the cosmic expansion history return $E(a)$ with some redshift-dependent uncertainty $\Delta E(a)$ \citep{2008A&A...481..295M, 2009A&A...508...45M, 2012MNRAS.419..513B, 2013MNRAS.436..854B, 2015JCAP...02..001T, 2018JCAP...11..027M, 2018IJMPA..3344022A, 2022ScPA....2....1H}. Given such an uncertainty, what uncertainty in $C_\ell^\gamma$ does it entail?

We use one such empirical reconstruction method for illustrating our results \citep{2022ScPA....2....1H}. It expands the suitably scaled, inverse expansion function into a set of orthonormal polynomials, specifically the shifted Chebyshev polynomials, and reconstructs the cosmic expansion function $E(a)$ by fitting this polynomial expansion to cosmic distance measurements, e.g.\ of type-Ia supernovae, or baryonic acoustic oscillations, or both in combination. The result of this fitting procedure is a small set of coefficients $c_i$ with uncertainties $\Delta c_i$. The measurement uncertainties quantified by $\Delta c_i$ imply an uncertainty $\Delta E(a)$ in the expansion function. Since the growth factor $D_+(a)$ of cosmic density fluctuations depends on the expansion function, any uncertainty in $E(a)$ will also cause an uncertainty $\Delta D_+(a)$ in the growth factor. Supplied with the specific functional form of $\Delta E(a)$ resulting from our reconstruction method, we wish to quantify how uncertain the weak-lensing power spectrum $C_\ell^\gamma$ is in response to the uncertainty in the empirically constrained cosmic expansion function.

Since our empirical reconstruction method returns the functional forms $E(a)$ and $\Delta E(a)$ of the expansion function and its uncertainty rather than a set of values of these functions at specific, discrete redshifts or scale factors, we require the functional derivative of the weak-lensing power spectrum $C_\ell^\gamma$ with respect to the expansion function $E(a)$. We will work this out in Sect.~2. In Sect.~3, we will then quantify the uncertainty of $C_\ell^\gamma$ with the specific example of $\Delta E(a)$ as derived earlier from type-Ia supernovae and baryonic acoustic oscillations. In Sect.~4, we will then summarise and discuss our results.

\section{Functional derivative of the shear power spectrum}
\label{sec:2}

\subsection{The shear power spectrum}
\label{sec:2.1}

The power spectrum of cosmological weak gravitational lensing, specifically the expectation for the power spectrum of the weak gravitational shear $\gamma$ caused by cosmic structures, can be expressed as \citep{2001PhR...340..291B, 2010CQGra..27w3001B, schneider2006gravitational}.
\begin{equation}
  C_\ell^{\gamma}(w_\mathrm{s}) = \frac{9}{4}\left(\frac{H_0}{c}\right)^4
 \Omega_{\mathrm{m0}}^2\int_{0}^{w_\mathrm{s}}\D w\left(
 \frac{w_\mathrm{s} - w}{w_\mathrm{s} a(w)}
 \right)^2 P_{\delta}\left(\frac{\ell}{w}\right)\;.
\label{eq:1}
\end{equation}
The shear power spectrum $C_\ell^\gamma$ at an angular wave number $\ell$ is determined by a line-of-sight integral of the density-fluctuation power spectrum $P_\delta(k)$ at wave number $k=\ell/w$, geometrically weighed with the squared distance ratio between the matter fluctuations acting as gravitational lenses and the sources. In (\ref{eq:1}), $w$ is the comoving radial distance running from the source at $w = w_\mathrm{s}$ to the observer at $w = 0$. The parameter of the present matter density is $\Omega_\mathrm{m0}$, and $cH_0^{-1}$ is the Hubble radius. Since density fluctuations grow with time, and thus also grow along the line-of-sight, variations of the growth factor change the density-fluctuation power spectrum $P_\delta$ and thus also affect our expectation for the shear power spectrum. In addition to the indirect variation mediated by the growth factor, the shear power spectrum is also directly affected by variations in the expansion function because the cosmic expansion affects the geometry of space and thereby all distance measures.

We substitute the comoving distance $w$ in (\ref{eq:1}) by the scale factor $a$ as an integration variable and find
\begin{equation}
\begin{aligned}
  C_\ell^{\gamma}(a_\mathrm{s}) &= \frac{9}{4}\left(\frac{H_0}{c}\right)^4 
  \Omega_{\mathrm{m0}}^2 \\ &\quad \times \int_{a_\mathrm{s}}^{1}\left(
    \frac{w(a_\mathrm{s}) - w(a)}{w(a_\mathrm{s})}
  \right)^2 P_{\delta}\left(\frac{\ell}{w(a)}\right)\frac{\D a}{a^4 E(a)}\;.
  \end{aligned}
\label{eq:2}
\end{equation}
Aiming at analyzing the impact of variations in the expansion function $E(a)$ and the growth factor $D_+(a)$ on this shear power spectrum, we can restrict ourselves to the integral in (\ref{eq:2}) because the prefactors are constants. Thus, our main object of interest is the integral
\begin{equation}
  I(a_\mathrm{s},\ell) := \int_{a_\mathrm{s}}^1\mathcal{W}^2(a,a_\mathrm{s})\,
  P_{\delta}\left(\frac{\ell}{w(a)}\right)\frac{\D a}{a^4 E(a)}
\label{eq:3}
\end{equation}
containing the geometrical weight function
\begin{equation}
  \mathcal{W}(a,a_\mathrm{s}) := \frac{w(a_\mathrm{s}) - w(a)}{w(a_\mathrm{s})}\;.
\label{eq:4}
\end{equation}

\subsection{Variation with the expansion function}
\label{sec:2.2}

Our uncertainty analysis proceeds in the three steps enumerated below. Within each step, we consider different angular wave numbers $\ell$ and compare results for linearly and non-linearly evolving density-fluctuation power spectra $P_\delta$.
\begin{enumerate}
  \item As motivated in the introduction, we wish to answer the question: If we vary the cosmic expansion function by $\Delta E(x)$ at a certain scale factor $x$, which we henceforth call the \emph{perturbation scale factor}, how does the integral $I(a_\mathrm{s}, \ell)$ change relatively? Moving the perturbation scale factor $x$ from $a_\mathrm{s}$ to unity then results in a relative uncertainty distribution for fixed $a_\mathrm{s}$ and $\ell$, depending on $x$, which can be calculated using Gaussian error progression,
  \begin{equation}
    R(x, a_\mathrm{s}, \ell) := \frac{\Delta I_x(a_\mathrm{s},\ell)}{I(a_\mathrm{s},\ell)} =
    \left|\fdv{\ln I(a_\mathrm{s},\ell)}{E(x)}\right|\cdot\Delta E(x)\;.
  \label{eq:5}
  \end{equation}
  We shall refer to $R(x, a_\mathrm{s}, \ell)$ as the two-dimensional uncertainty distribution since we shall later fix $\ell$ and vary $x$ and $a_\mathrm{s}$. The function $R(x, a_\mathrm{s}, \ell)$ answers the question: What is the relative change of the weak-lensing power spectrum $C_\ell^\gamma$ in response to a change in the expansion function somewhere between the source and the observer?
  
  Variations of the expansion function $E(a)$ also cause variations of the growth factor $D_+(a)$. Possible additional systematic uncertainties in $D_+$ are negligible in comparison. Thus, the uncertainties $\Delta D_+$ are dominated by the uncertainties $\Delta E$, propagated through the differential equation (\ref{eq:18}) for linear density fluctuations. Thus, $\Delta D_+$ depends linearly on $\Delta E$ with the functional derivative $\delta D_+/\delta E$ as the slope, which will be calculated in Sect.~\ref{sec:2.2.3}. Thus, the uncertainties in $D_+$ are implicitly taken into account in (\ref{eq:5}), as we shall also see in Sect.~\ref{sec:2.2}.
 
  \item We additionally vary the scale factor $a_\mathrm{s}$ of the source and thus arrive at the two-dimensional relative uncertainty distribution $R(x, a_\mathrm{s}, \ell)$ at fixed $\ell$. We restrict our analysis to sources at scale factors $a_\mathrm{s}\in[1/3,2/3]$, corresponding to source redshifts $z_\mathrm{s}\in[0.5, 2]$.
  
  \item Accounting for all perturbations along the line-of-sight, achieved by integrating the relative uncertainty distribution over all perturbation scale factors $x$, finally gives the total relative uncertainty for a given source at scale factor $a_\mathrm{s}$ as a function of the angular wave number $\ell$,
  \begin{equation}
    \bar R(a_\mathrm{s}, \ell) = \int_{a_\mathrm{s}}^1\D x\,R(x, a_\mathrm{s}, \ell) =
    \int_{a_\mathrm{s}}^1\D x\,\frac{\Delta I_x(a_\mathrm{s},\ell)}{I(a_\mathrm{s},\ell)}\;.
  \label{eq:6}
  \end{equation}
  We shall refer to this function as one-dimensional since we vary $a_\mathrm{s}$ at fixed angular scale $\ell$.
\end{enumerate}

\subsubsection{Variation of the geometrical weight function}
\label{sec:2.2.1}

To quantify the relative variation $R(x, a_\mathrm{s}, \ell)$ defined in (\ref{eq:5}), we need the functional derivative of $I(a_\mathrm{s}, \ell)$ with respect to the expansion function $E(x)$. Applying the rules of functional derivation, we have
\begin{align}
  &\fdv{I(a_\mathrm{s},\ell)}{E(x)} \nonumber \\ 
  &= \int_{a_\mathrm{s}}^{1}\frac{\D a}{a^4 E(a)}
    \Theta(x-a_\mathrm{s})\Theta(1-x)
    \Bigg\{
    \mathcal{W}^2(a,a_\mathrm{s}) P_{\delta}\left(\frac{\ell}{w(a)},D_+(a)\right)
  \nonumber\\ & \quad \times
  \left[
    2\fdv{\ln\mathcal{W}(a,a_\mathrm{s})}{E(x)}+
    \fdv{\ln P_{\delta}\left(\ell/w(a), D_+(a)\right)}{E(x)}-
    \frac{\delta(x-a)}{E(a)}
  \right]\Bigg\}\;.
\label{eq:7}
\end{align}
The functional derivative of the geometrical weight function $\mathcal{W}$ with respect to $E$ is
\begin{equation}
  \fdv{\mathcal{W}(a,a_\mathrm{s})}{E(x)} = \left[
    \fdv{w(a_\mathrm{s})}{E(x)} w(a) -\fdv{w(a)}{E(x)} w(a_\mathrm{s})
  \right]\frac{1}{w^2(a_\mathrm{s})}\;,
\label{eq:8}
\end{equation}
and the functional derivative of the comoving radial distance with respect to the expansion function turns out to be
\begin{align}
  \fdv{w(a)}{E(x)} &= \fdv{}{E(x)}\int_{a}^{1}\frac{\D a'}{a'^2 E(a')} \nonumber \\
  &= -\int_{a}^{1}\frac{\D a'}{a'^2}\,
  \frac{\delta_D(a' - x)}{E^2(a)}\,
  \Theta(x - a)\Theta(1 - x) \nonumber\\ &=
  -\frac{\Theta(x - a)\Theta(1 - x)}{x^2 E^2(x)}\;.
\label{eq:9}
\end{align}
Combining (\ref{eq:9}) with (\ref{eq:8}) and using that
\begin{equation}
  \Theta(x-a) = \Theta(x-a_\mathrm{s})\left[1 -\Theta(a-x)\right]
\label{eq:10}
\end{equation}
for $a<a_\mathrm{s}$ leads to the compact result 
\begin{equation}
  \fdv{\ln\mathcal{W}(a,a_\mathrm{s})}{E(x)} =
  \frac{\Theta(1-x)\Theta(x-a_\mathrm{s})}{w(a_\mathrm{s}) x^2 E^2(x)}\left[
    1 -\frac{\Theta(a-x)}{\mathcal{W}(a,a_\mathrm{s})}
  \right]\;.
\label{eq:11}
\end{equation}

\subsubsection{Variation of the power spectrum}
\label{sec:2.2.2}

The functional derivative of the density-fluctuation power spectrum $P_\delta$ due to changes in the expansion function can be calculated as follows:

\begin{align}
  \fdv{\ln P_{\delta}\left(k,D_+(a)\right)}{E(x)}  &=
  \pdv{\ln P_{\delta}}{k}\pdv{k}{w(a)}\fdv{w(a)}{E(x)}+
  \pdv{\ln P_{\delta}}{D_+}\fdv{D_+(a)}{E(x)} \nonumber \\ &=
  \frac{\Theta(1-x)\Theta(x-a)}{w^2(a)x^2E^2(x)}
  \pdv{\ln P_{\delta}}{\ln k}\bigg|_{k = \ell/w(a)} \nonumber \\ & \quad +
  \pdv{\ln P_{\delta}}{D_+}\fdv{D_+(a)}{E(x)}\;. 
\label{eq:12}
\end{align}

The linearly evolving power spectrum satisfies $P_{\delta}^\mathrm{(lin)} = D_+^2 P_{\delta0}^\mathrm{(lin)}$, such that
\begin{equation}
  \pdv{\ln P_{\delta}^\mathrm{(lin)}}{D_+} = \frac{2}{D_+}\;,
\label{eq:13}
\end{equation}
whereas the dependence of the non-linearly evolving power spectrum $P_\delta^\mathrm{(nl)}$ on $D_+$ is given implicitly by
\begin{align}
  \pdv{\ln P_{\delta}^\mathrm{(nl)}}{D_+} &=
  \left(\dv{D_+}{a}\right)^{-1}\pdv{\ln P_{\delta}^\mathrm{(nl)}}{a} =
  \left(\frac{D_+}{a}\dv{\ln D_+}{\ln a}\right)^{-1}
  \pdv{\ln P_{\delta}^\mathrm{(nl)}}{a} \nonumber\\ &=
  \frac{a}{D_+\Omega_\mathrm{m}^\gamma}\pdv{\ln P_{\delta}^\mathrm{(nl)}}{a} =
  \frac{1}{D_+\Omega_\mathrm{m}^\gamma}
  \pdv{\ln P_{\delta}^\mathrm{(nl)}}{\ln a}\;.
\label{eq:14}
\end{align}

\begin{figure}
  \centerline{\includegraphics[width=\hsize]{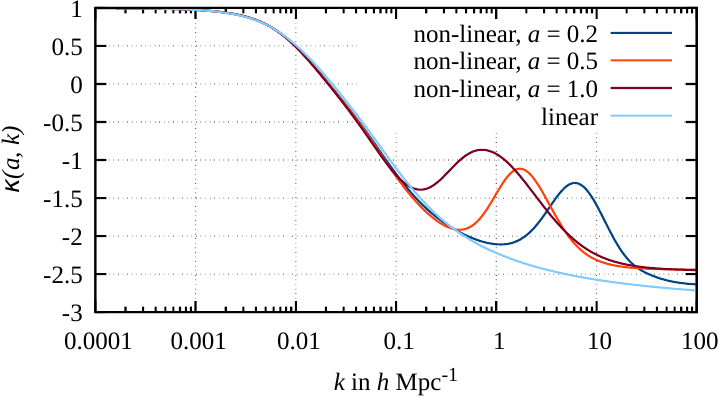}}
\caption{The logarithmic derivative $\kappa := \partial\ln P_\delta/\partial\ln k$ defined in (\ref{eq:16}) is shown here for linear and non-linear density-fluctuation power spectra and for scale factors $a = 0.2, 0.5, 1.0$. For small $k$, the derivative tends to unity. The peak at small $k$ in the derivatives of the non-linear power spectra marks the scale below which non-linearity sets in.}
\label{fig:1}
\end{figure}

To combine both cases, i.e.\ the linearly and the non-linearly evolving density-fluctuation power spectra, we introduce
\begin{equation}
  \alpha(k,a) := \pdv{\ln P_{\delta}(k,a)}{\ln D_+} = 
  \begin{cases}
    2 & \mbox{(linear)} \\
    \Omega_\mathrm{m}^{-\gamma}(a)\pdv{\ln P_{\delta}(k,a)}{\ln a} &
      \mbox{(non-linear)}
  \end{cases}\;.
\label{eq:15}
\end{equation}
Here, $\gamma$ is implicitly given by the logarithmic derivative
\begin{equation}
  \frac{\D\ln D_+}{\D\ln a} = \Omega^\gamma_\mathrm{m}(a)
\label{eq:15a}
\end{equation}
of the growth factor with respect to the scale factor. Further defining the logarithmic derivative of both types of density-fluctuation power spectra with respect to the wave number by
\begin{equation}
  \kappa(k,a) := \pdv{\ln P_{\delta}(k,a)}{\ln k}\;,
\label{eq:16}
\end{equation}
we can summarise the functional derivative of both types of power spectra by
\begin{equation}
\begin{aligned}
  \fdv{\ln P_{\delta}\left[k,D_+(a)\right]}{E(x)} &=
  \frac{\Theta(1-x)\Theta(x-a)}{w(a) x^2 E^2(x)}\,\kappa(k,a) \\ & \quad +
  \fdv{\ln D_+(a)}{E(x)}\,\alpha(k,a)\;.
  \end{aligned}
\label{eq:17}
\end{equation}
The first term on the right-hand side reflects the variation of the density-fluctuation power spectrum $P_\delta(k,a)$ in response to a change in the wave number $k$ where it is to be evaluated, which is in turn due to a change in the comoving radial distance $w(a)$ at fixed angular wave number $\ell$. {{The function $\kappa(k,a)$ defined in (\ref{eq:16}) also takes the shape evolution of the non-linear power spectrum into account; see Fig.~\ref{fig:1}}}. The radial comoving distance $w(a)$ itself responds to variations in the expansion function $E(x)$ with $x > a$, as given by (\ref{eq:9}). The second term on the right-hand side reflects the variation of $P_\delta(k,a)$ with uncertainties in the time evolution of cosmic structures in response to changes in the expansion function $E(x)$.

\begin{figure}
\centering
  \includegraphics[width=\hsize]{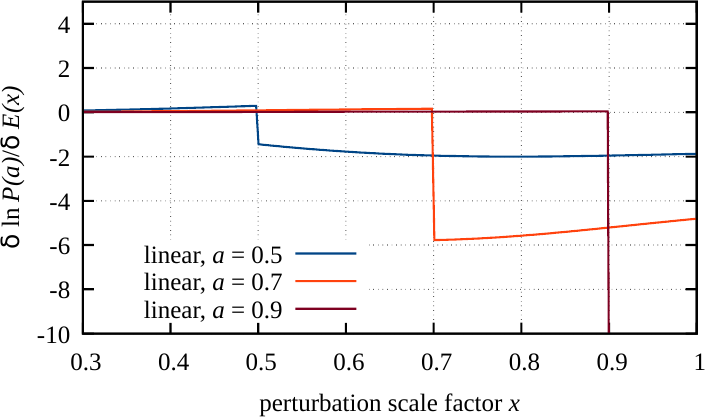} 
  \includegraphics[width=\hsize]{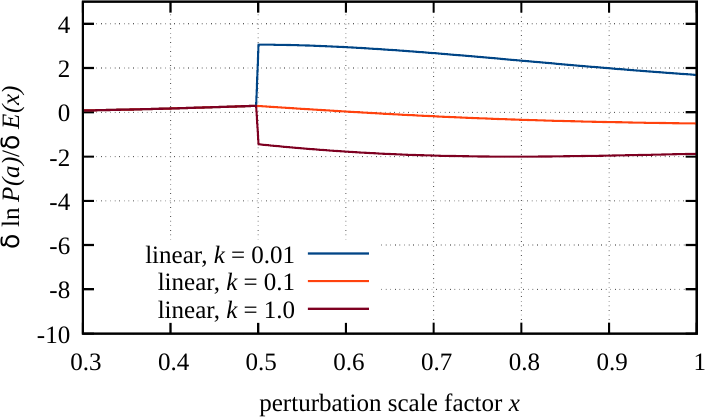}
\caption{Logarithmic derivative of the linear density-fluctuation power spectrum with respect to the expansion function, $\delta P_\delta(k,a)/\delta E(x)$. \emph{Upper panel}: derivatives taken at $k = 1.0\,h\,\mathrm{Mpc}^{-1}$ and scale factors $a = 0.5,0.7,0.9$. \emph{Lower panel}: derivatives taken at $a = 0.5$ and wave numbers $k = 0.01,0.1,1.0\,h\,\mathrm{Mpc}^{-1}$.}
\label{fig:2a}
\end{figure}

\begin{figure}
    \centering
    \includegraphics[width=\hsize]{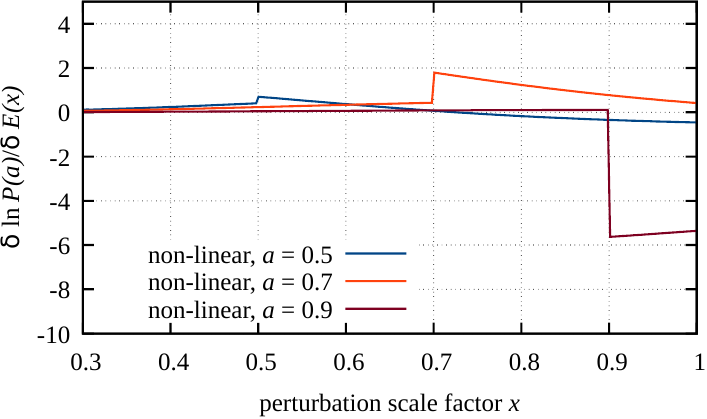}  
    \includegraphics[width=\hsize]{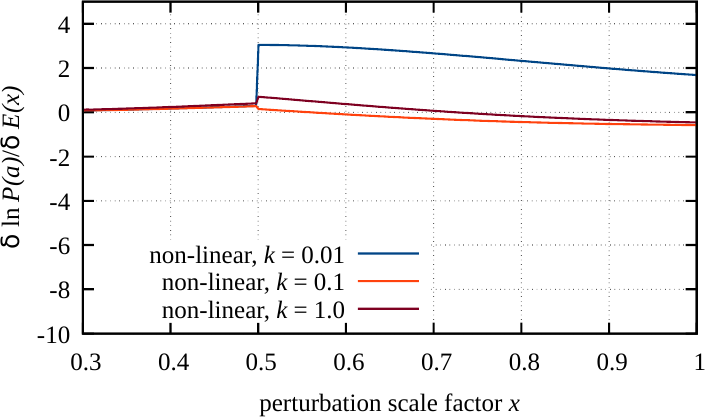}
    \caption{Logarithmic derivative of the non-linear density-fluctuation power spectrum with respect to the expansion function, $\delta P_\delta(k,a)/\delta E(x)$. \emph{Upper panel}: derivatives taken at $k = 1.0\,h\,\mathrm{Mpc}^{-1}$ and scale factors $a = 0.5,0.7,0.9$. \emph{Lower panel}: derivatives taken at $a = 0.5$ and wave numbers $k = 0.01,0.1,1.0\,h\,\mathrm{Mpc}^{-1}$.}
    \label{fig:2b}
\end{figure}

For evaluating the functional derivative (\ref{eq:17}), we use the linearly evolved CDM power spectrum by \citep{1986ApJ...304...15B} and the non-linearly evolved CDM power spectrum by \citep{2003MNRAS.341.1311S}. There are clearly newer descriptions of the non-linear density-fluctuation power spectrum, but we have convinced ourselves that its detailed shape is unimportant for our purposes. The function $\kappa$ is shown in Fig.~\ref{fig:1}, the functional derivatives of the power spectra with respect to the expansion function are illustrated in Fig.~\ref{fig:2a} (linear case) and Fig.~\ref{fig:2b} (non-linear case).

The discontinuity of the functional derivative $\delta\ln P_\delta(a)/\delta E(x)$ at $x = a$ is the combined effect of the two terms in (\ref{eq:17}). While the first term occurs only for $x>a$ because it is due to varying the radial comoving distance from the observer to the density fluctuation, the second term also contributes at $x<a$ because it reflects the growth of density fluctuations over time.

\subsubsection{Variation of the growth factor}
\label{sec:2.2.3}

The final piece required for calculating the relative variation $R(x, a_\mathrm{s}, \ell)$ is the functional derivative of the growth factor with respect to the expansion function, $\delta D_+(a)/\delta E(x)$. Both functions are related via the differential equation
\begin{equation}
  D_+''(a)+\left[\frac{3}{a}+\frac{E'(a)}{E(a)}\right]D_+'(a)-
  \frac{3}{2}\frac{\Omega_\mathrm{m}(a)}{a^2} D_+(a) = 0\;,
\label{eq:18}
\end{equation}
where the prime denotes the derivative with respect to the scale factor $a$. The matter-density parameter $\Omega_\mathrm{m}(a)$ as a function of the scale factor is given by
\begin{equation}
  \Omega_\mathrm{m}(a) = \frac{\Omega_{\mathrm{m}0}}{a^3E^2(a)}
\label{eq:19}
\end{equation}
in terms of the expansion function.

A straightforward approach at the desired functional derivative begins with applying the functional derivative $\delta/\delta E(x)$ to the differential equation (\ref{eq:18}) and making use of the fact that the functional derivative $\delta/\delta E(x)$ commutes with the derivative $\D/\D a$ with respect to the scale factor. Doing so, we arrive at a differential equation for $\delta D_+(a)/\delta E(x)$,
\begin{equation}
  \fdv{}{E(x)}\left[
    D_+''(a)+\left(
      \frac{3}{a} +\frac{E'(a)}{E(a)}
    \right)D_+'(a) -\frac{3}{2}\frac{\Omega_\mathrm{m}(a)}{a^2} D_+(a)
  \right] = 0\;.
\label{eq:20}
\end{equation}
Carrying out the functional derivative results in
\begin{align}
  &\dvtw{}{a}\fdv{D_+(a)}{E(x)}+\left(
    \frac{3}{a}+\dv{\ln E(a)}{a}
  \right)\dv{}{a}\fdv{D_+(a)}{E(x)}-
  \frac{3}{2}\frac{\Omega_\mathrm{m}(a)}{a^2}\fdv{D_+(a)}{E(x)} \nonumber \\ &=
  -\dv{}{a}\left(
    \frac{\delta_D(a-x)}{E(a)}
  \right)D_+'(a)-3\frac{D_+(a)}{a^2}\frac{\Omega_\mathrm{m}(a)}{E(a)}\delta_D(a-x)\;.
\label{eq:21}
\end{align}
Recognizing that the part of (\ref{eq:21}) which is homogeneous in the functional derivative is of the same form as (\ref{eq:18}), which is solved by $D_+(a)$, the ansatz
\begin{equation}
  \fdv{D_+(a)}{E(x)} = C(a,x)\cdot D_+(a)
\label{eq:22}
\end{equation}
suggests itself. Moreover, we enforce causality explicitly on the solution of (\ref{eq:21}) by requiring that a perturbation in $E(x)$ at $x > a$ cannot propagate backwards in time to disturb $D_+(a)$. We thus impose
\begin{equation}
  \fdv{D_+(a)}{E(x)}\propto \Theta(a-x)\;,
\label{eq:23}
\end{equation}
which fixes the two degrees of freedom of this second-order differential equation, such that we do not need to set any further boundary conditions. Using the ansatz (\ref{eq:22}) and the causality condition (\ref{eq:20}) leads to the solution
\begin{equation}
  \fdv{D_+(a)}{E(x)} = D_+(a) \Theta(a-x) [g(x)\Gamma(x, a) - h(x)] \;,
\label{eq:24}
\end{equation}
with the functions $g(x),h(x)$ and $\Gamma(a,x)$ defined by
\begin{equation}
\begin{aligned}
  g(x) &= x D_+^2(x)\Omega_\mathrm{m}(x)
  \left[\Omega_\mathrm{m}^{2\gamma - 1}(x)-\frac{3}{2}\right]\;,\\ 
  h(x) &= \frac{\Omega_{\mathrm{m}(x)}}{x E(x)}\;,\\
  \Gamma(x, a) &= \int_{x}^{a}\frac{\D a'}{a'^3 D_+^2(a') E(a')}\;.
  \end{aligned}
\label{eq:25}
\end{equation}

Finally, we need to take into account that the density-fluctuation power spectrum is normalised at the present time. This means that, by the choice of this normalisation, the growth factor $D_+$ has to be set to unity today, $D_+(a = 1) = 1$. The variation of the growth factor normalised in this way is 
\begin{equation}
\begin{aligned}
  \frac{\delta}{\delta E(x)}\left(\frac{D_+(a)}{D_+(1)}\right) &=
  \frac{1}{D_+(1)}\frac{\delta D_+(a)}{\delta E(x)}-
  \frac{D_+(a)}{D_+^2(1)}\frac{\delta D_+(1)}{\delta E(x)} \\
  &=
  \frac{\delta D_+(a)}{\delta E(x)}-D_+(a)\frac{\delta D_+(1)}{\delta E(x)}\;,
  \end{aligned}
\label{eq:27}
\end{equation} 
replacing $D_+(1)$ by unity in the second step. With (\ref{eq:24}), we can write
\begin{equation}
  \frac{\delta}{\delta E(x)}\left(\frac{D_+(a)}{D_+(1)}\right) =
  D_+(a) \{h(x)\Theta(x-a) - g(x) \Gamma[\text{max}(x,a),1] \} \;. 
\label{eq:28}
\end{equation}
Inserting (\ref{eq:11}), (\ref{eq:17}), and (\ref{eq:28}) into (\ref{eq:7}), we are finally in a position to work out the functional derivative of $I(a_\mathrm{s},\ell)$ with $E(a)$ numerically.

\section{Numerical evaluation and results}
\label{sec:3}

\subsection{General procedure and test with Einstein-de Sitter}
\label{sec:3.1}

This numerical evaluation needs the following inputs:
\begin{itemize}
  \item the expansion function $E(a)$ and its derivative as well as the growth factor $D_+$;
  \item the power spectrum $P_\delta$ of matter-density fluctuations;
  \item the matter-density parameter today, $\Omega_{\mathrm{m0}}$; and
  \item the exponent $\gamma$ of the logarithmic derivative of the growth factor.
\end{itemize}
With these ingredients, we can evaluate the functional derivative $\delta D_+(a)/\delta E(x)$ solving (\ref{eq:24}) with standard integration routines, in our case taken from the GNU Science Library. The derivatives $\alpha(k,a)$ and $\kappa(k,a)$ of the power spectrum with respect to the scale factor $a$ and the wave number $k$, defined in (\ref{eq:15}) and (\ref{eq:16}), also need to be calculated numerically except for the derivative of the linear power spectrum with respect to $a$.

\begin{figure}
\centering
  \includegraphics[width=0.9\hsize]{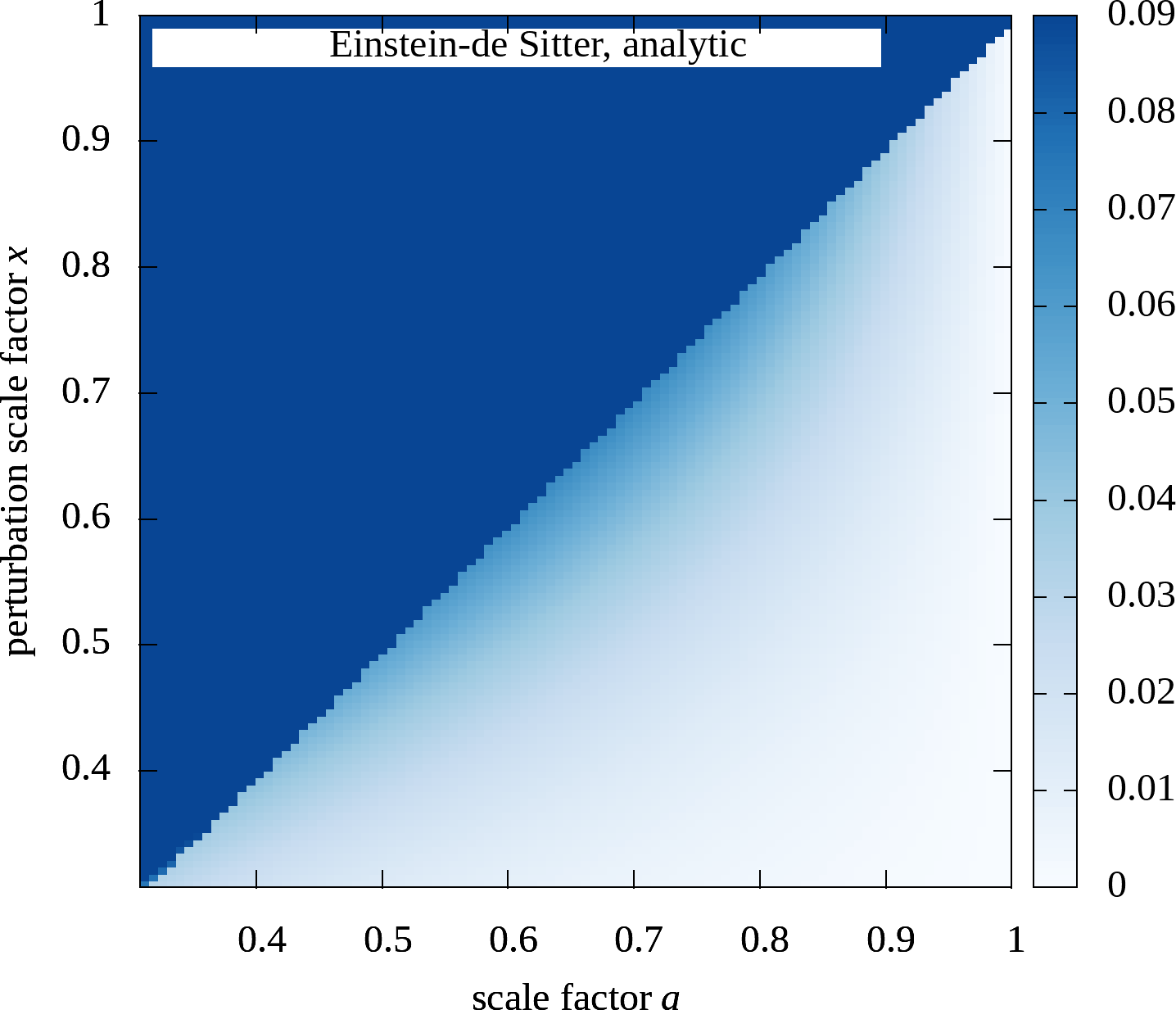} 
  \includegraphics[width=0.9\hsize]{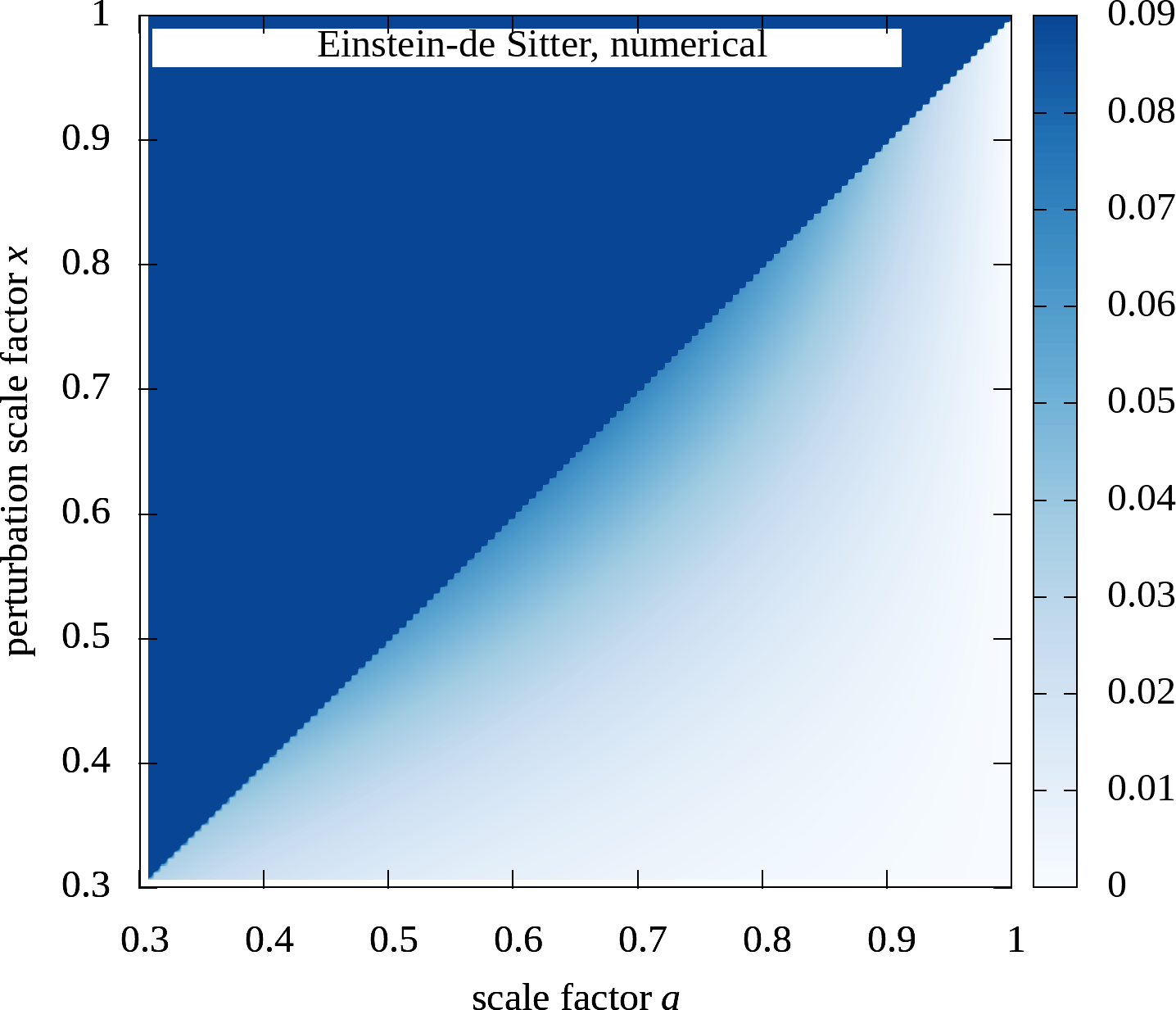}
  \includegraphics[width=0.9\hsize]{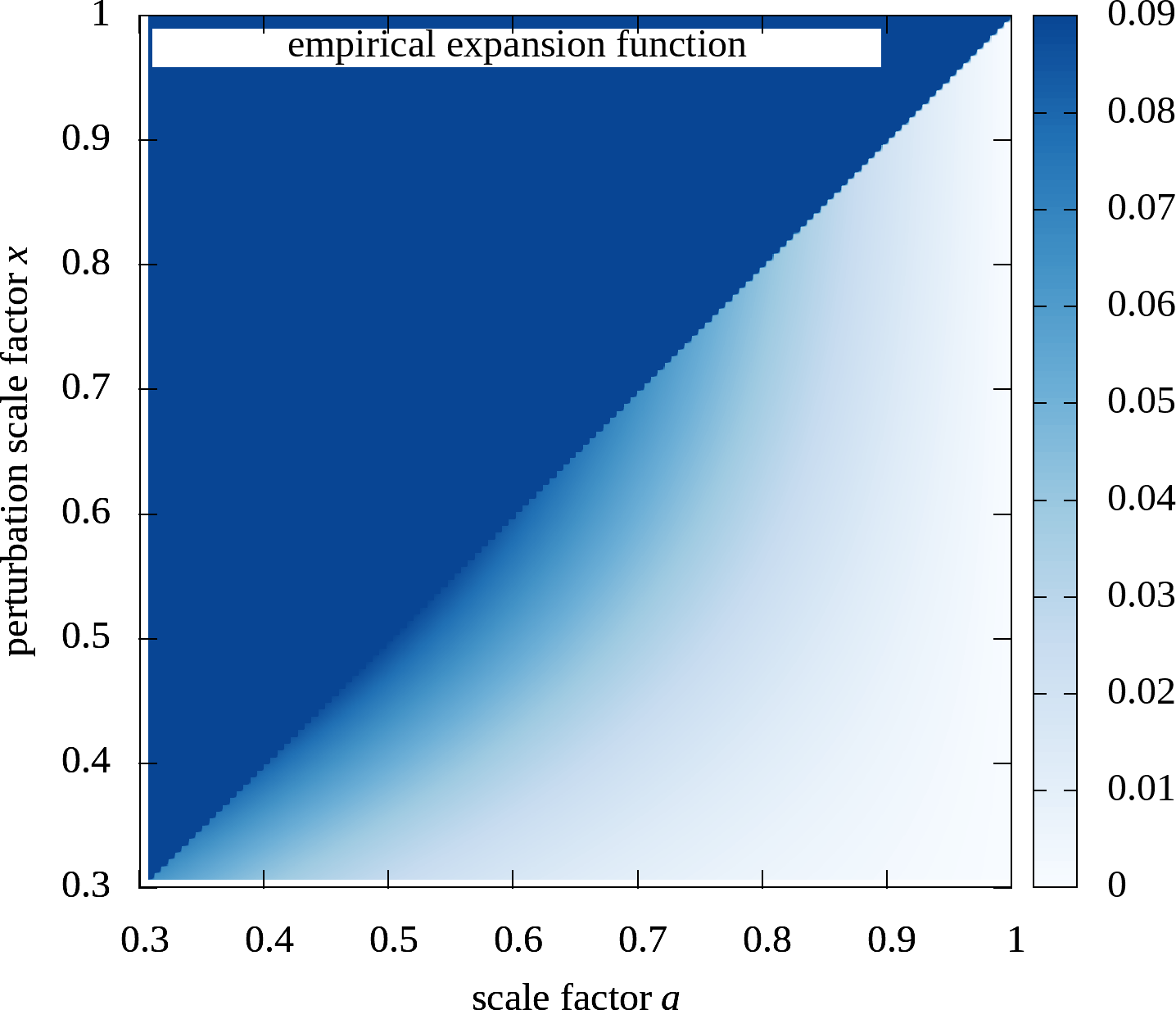}
\caption{Functional derivative of the normalised growth factor with respect to the expansion function for two different cases. Upper panel: analytic result for an Einstein-de Sitter universe for reference; central panel: numerical result for an Einstein-de Sitter universe;  lower panel: numerical result obtained using an empirically constrained expansion function.}
\label{fig:3}																												%
\end{figure}

It is interesting to consider an Einstein-de Sitter universe as a reference cosmology. It contains only matter with critical density. Since all the input functions are then known analytically, we can compare the analytic solution for the functional derivative in such a model universe with the numerical results. The characteristic functions in this universe are 
\begin{equation}
  E(a) = a^{-3/2}\;,\quad D_+(a) = a\;,\quad \Omega_\mathrm{m}(a) = 1\;.
\label{eq:29}
\end{equation}
Using these, we find from (\ref{eq:25})
\begin{equation}
  g(x) = -\frac{x^3}{2}\;, \quad h(x) = x^{1/2}\;, \quad \Gamma(x, a) = -\frac{2}{5}\left(
    a^{-5/2}-x^{-5/2}
  \right)
\label{eq:30}
\end{equation} 
for the functions $g(x),h(x)$ and $\Gamma(x,a)$ and, from (\ref{eq:28}), the functional derivative of the normalised growth factor becomes
\begin{equation}
  \fdv{}{E(x)}\left(\frac{D_+(a)}{D_+(1)}\right) = a \bigg\{  
  \frac{x^3}{5} \left[{\max}^{-5/2}(x, a) - 1\right] + \sqrt{x} \Theta(x-a)\bigg\} \;.
\label{eq:31}
\end{equation}
This analytic result is shown in the top panel of Fig.~\ref{fig:3}. It is opposed there to the numerical result for an Einstein-de Sitter universe in the centre panel for comparison, and to the functional derivative of $D_+$ obtained by evaluating (\ref{eq:28}) numerically as described in the following subsection, shown in the bottom panel. The analytical and the numerical results for the Einstein-de Sitter universe are identical, thus confirming our numerical implementation. Despite the difference in the cosmological background evolution between the left and central panels compared to the right panel, the results are quite similar.

\subsection{Evaluation with empirical expansion function}
\label{sec:3.2}

\subsubsection{Empirical expansion function}
\label{sec:3.2.1}

\begin{figure}
\centering
  \centerline{\includegraphics[width=\hsize]{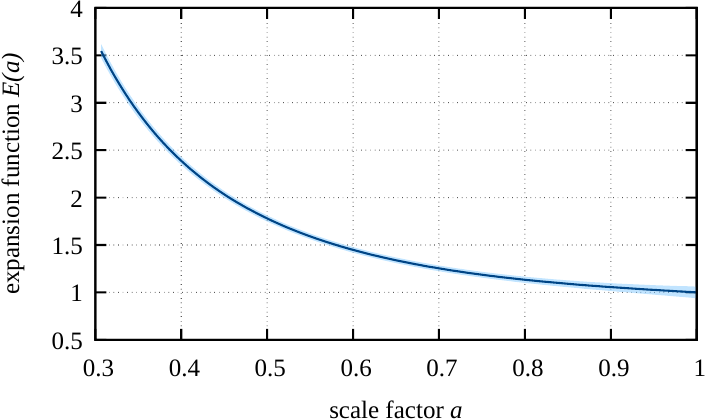}}
\caption{The model independent expansion function with its uncertainty}
\label{fig:4}
\end{figure}

Our model-independent reconstruction of the expansion function, described in detail elsewhere \citep{2022ScPA....2....1H}, proceeds as follows. Given measurements of luminosity distances (or distance moduli) with redshifts $z\le z_\mathrm{max}$, we define the re-scaled scale factor $x$ by   
\begin{equation}
  x = \frac{a-a_\mathrm{min}}{1-a_\mathrm{min}}\;,\quad a_\mathrm{min} =
  \left(1+z_\mathrm{max}\right)^{-1}\;.
\label{eq:32}
\end{equation} 
With the convenient further definitions
\begin{equation}
  \delta a := \frac{1-a_\mathrm{min}}{a_\mathrm{min}}\;,\quad
  e(x) := \frac{H_0}{\dot x(1+x\delta a)}
\label{eq:33}
\end{equation}
we can write the luminosity distance in a spatially-flat universe in the form
\begin{equation}
  D_\mathrm{lum}(x) = \frac{c}{H_0}\frac{1}{a_\mathrm{min}^2(1+x\delta a)}
  \int_x^1\D x'\,e(x')\;.
\label{eq:34}
\end{equation}
We expand $e(x)$ in a series of shifted Chebyshev polynomials $T_j^*(x)$
\begin{equation}
  e(x) = \sum_{j=0}^Mc_jT_j^*(x)\;,
\label{eq:35}
\end{equation}
and determine the coefficients $c_j$ by a maximum-likelihood fit of $D_\mathrm{lum}(x)$ to the measurements. Their uncertainties $\Delta c_j$, and thus the uncertainty $\Delta e(x)$ of $e(x)$, are determined by the covariance matrix of the data. The expansion function is
\begin{equation}
  E(x) = \frac{\delta a}{(1+x\delta a)^2e(x)}\;,
\label{eq:36}
\end{equation}
in terms of $e(x)$, and its uncertainty follows from
\begin{equation}
  \left|\frac{\Delta E(x)}{E(x)}\right| =
  \left|\frac{\Delta e(x)}{e(x)}\right|\;.
\label{eq:37}
\end{equation}
The expansion function derived in this way from type-Ia supernovae and BAO data is shown together with its uncertainty in Fig.~\ref{fig:4}. 

Since the expansion function $E(a)$ is given in terms of Chebyshev polynomials, its derivative with respect to the scale factor $a$ can be taken. We define
\begin{equation}
  \varepsilon := 3+2\,\frac{\D\ln E(a)}{\D\ln a}\;,\quad
  \omega := 1-\Omega_\mathrm{m}(a)
\label{eq:38}
\end{equation}
where $\Omega_\mathrm{m}(a)$ is determined by (\ref{eq:19}) in terms of the expansion function for any choice of the matter-density parameter $\Omega_\mathrm{m0}$ today. With $\varepsilon$ and $\omega$ from (\ref{eq:38}), the growth index $\gamma$ is given by \citep{1998ApJ...508..483W, 2017PhRvD..95b3505A, 2022ScPA....2....1H}
\begin{equation}
  \gamma = \frac{\varepsilon+3\omega}{2\varepsilon+5\omega}\;.  
\label{eq:39}
\end{equation}
Thus, with $E(a)$ and $\Delta E(a)$ reconstructed from measurements, we only need to set $\Omega_\mathrm{m0}$. We choose $\Omega_\mathrm{m0} = 0.3$.

\subsubsection{Functional derivative of the growth factor}
\label{sec:3.2.2}

The model-independent expansion function and growth factor are available in the scale-factor range $0.307\le a\le 0.990$. We restrict ourselves to $a \in [1/3,1]$ because the later uncertainty analysis considers source scale factors $a_\mathrm{s}$ in this range. Numerically computing the functional derivative $\delta D_+(a)/\delta E(x)$ using the empirically determined expansion function and its uncertainty gives the result shown in the right panel of Fig.~\ref{fig:3}.

As for the Einstein-de Sitter result, the functional derivative of the normalised growth factor is positive on the entire domain. This illustrates that, with the growth factor fixed to unity today, an increase in the expansion function causes an increase in structure growth: if structures are to reach their present amplitude in a more rapidly expanding universe, they need to grow faster against the more rapidly expanding background. In the Einstein-de Sitter universe, structure growth is somewhat delayed compared to our actual universe, which expands more rapidly than Einstein-de Sitter at late times. Once more, this reflects the fact that structures need to grow earlier in a universe expanding more quickly if their present amplitude is fixed.

\subsection{Evaluation of the uncertainty distributions}
\label{sec:3.3}

\begin{figure}
  \centerline{\includegraphics[width=\hsize]{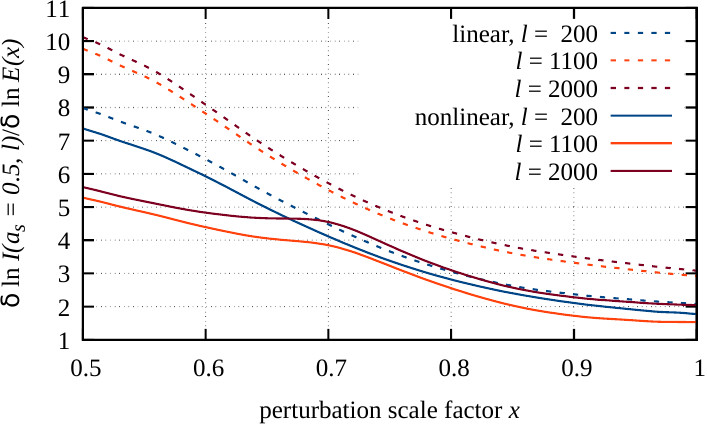}}
\caption{The figure shows the logarithmic derivative $\delta\ln I(a_\mathrm{s}, \ell)/\delta\ln E(x)$ of the shear power spectrum with respect to the expansion function. The source scale factor is set to $a_\mathrm{s} = 0.5$ here, the angular wave numbers are chosen from $\ell\in[200, 1100, 2000]$, as indicated. The dashed and solid lines show results for the linearly and the non-linearly evolving density-fluction power spectrum, respectively. For non-linearly growing density perturbations, the logarithmic derivative is smaller because the cosmic-shear signal is substantially enhanced by non-linear compared to linear density fluctuations. Moreover, for non-linear density fluctuations, the logarithmic derivative flattens at perturbation scale factors of $\approx 0.7-0.8$ for intermediate and small angular wave numbers. This is due to the peak in the logarithmic derivative $\kappa(a, k)$ of the density-fluctuation power spectrum caused by the onset of non-linear structure growth.} 
\label{fig:7}
\end{figure}

With the functional derivative $\delta D_+(a)/\delta E(x)$ of the growth factor with respect to the empirically reconstructed expansion function at hand, we can now numerically evaluate the functional derivative of the integral $I(a_\mathrm{s}, \ell)$ from (\ref{eq:3}) with respect to the expansion function $E(a)$ and the uncertainty distributions $R(x, a_\mathrm{s}, \ell)$ and $\bar R(a_\mathrm{s}, \ell)$ defined in (\ref{eq:5}) and (\ref{eq:6}) based on the empirical expansion function $E(a)$ and its uncertainty $\Delta E(a)$. Note that $R(x, a_\mathrm{s}, \ell)$ differs from the logarithmic derivative $\delta\ln I(a_\mathrm{s}, \ell)/\delta E(x)$ by the factor $\Delta E(x)$, i.e.\ by a specific function quantifying the uncertainty of the expansion function. Thus, before we proceed to the uncertainty distributions $R$ and $\bar R$, we first show in Fig.~\ref{fig:7} the logarithmic derivative $\delta\ln I(a_\mathrm{s}, \ell)/\delta E(x)$ of the weak-lensing power spectrum with respect to the expansion function for sources at $a_\mathrm{s} = 0.5$ or $z = 1$ and for the three different angular wave numbers $\ell\in[200, 1100, 2000]$, corresponding to angular scales of $\theta \approx 180^\circ/\ell \in [54', 9.8', 5.4']$.

The figure illustrates the two interesting aspects that the logarithmic derivative of the shear power spectrum is larger for the linearly than for the non-linearly growing power spectrum, and that it flattens at perturbation scale factors $x\approx 0.7-0.8$ (i.e.\ at perturbation redshifts $z\approx 0.45-0.25$) as the angular scale shrinks, i.e.\ as $\ell$ increases. Both aspects are due to the non-linear growth, with substantially enhances the shear power spectrum and introduces a peak in the logarithmic derivative $\kappa(a, k)$ of the density-fluctuation power spectrum with respect to wave number, as shown in Fig.~\ref{fig:1}. For non-linearly growing structures, the functional derivatives vary little with both the perturbation scale factor and the angular wave number. On intermediate and small angular scales, it is {{$\approx 2-6$}}, showing that a relative change in the expansion function of, say, $1\,\%$ changes the shear power spectrum by {{$\approx2-6\,\%$.}}

\begin{figure}
\centering
  \includegraphics[width=0.8\hsize]{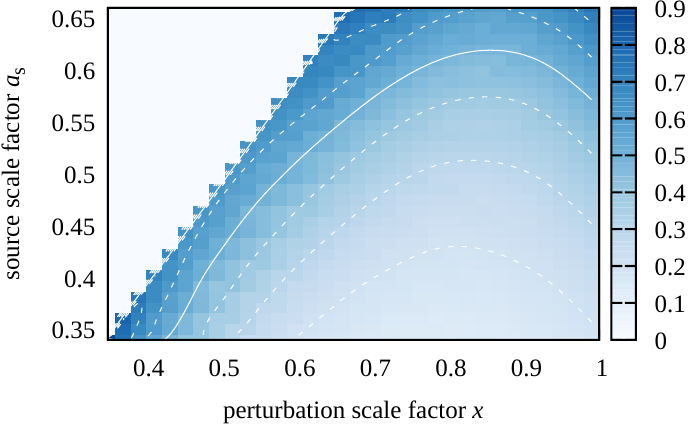}  
  \includegraphics[width=0.8\hsize]{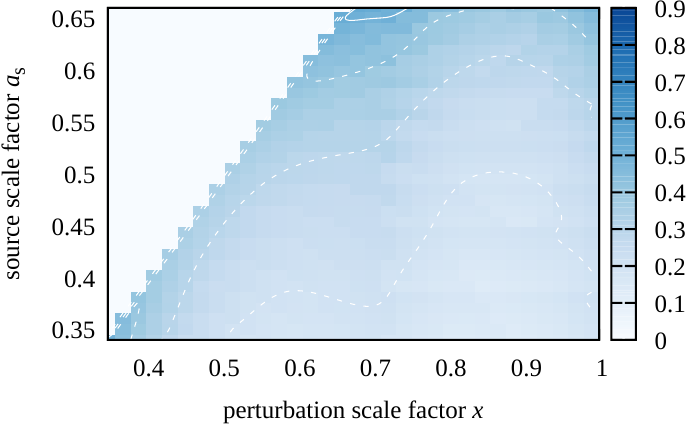}
\caption{The two-dimensional, relative uncertainty distribution $R(x, a_\mathrm{s}, \ell)$ as defined in (\ref{eq:5}) is shown here for $\ell = 2000$, approximately corresponding to an angular scale of $5.4'$. \emph{Upper Panel}: result for a linearly growing power spectrum; \emph{lower panel}: result for a non-linearly growing power spectrum. For $x < a_\mathrm{s}$, the function vanishes identically. The level of the heavy, solid contour is $0.5$; the dotted contours are spaced by $0.1$.}
\label{fig:5}
\end{figure}

We then proceed to the uncertainty distribution $R(x, a_\mathrm{s}, \ell)$ for $\ell = 2000 \approxcorr 5.4'$, which we show in Fig.~\ref{fig:5}. We focus on small angular scales because non-linear effects are important there \citep{2001PhR...340..291B, schneider2006gravitational}.

The two panels of Fig.~\ref{fig:5} show results assuming a linearly (top) and a non-linearly (bottom) evolving power spectrum. For $x < a_\mathrm{s}$, the distributions vanish identically because then the perturbation of the expansion function precedes the emission by the source. 

For linear structure growth, the distribution has a broad and shallow minimum at intermediate perturbation scale factors $x$ and grows for later-time perturbations, $x\to1$, and with decreasing source distance, $a_\mathrm{s}\to2/3$. Late-time perturbations affect cosmic distances and structure growth, and for less distant sources, the shear signal decreases. Both effects together cause the increase of $R$ for increasing source and perturbation scale factors. Non-linear structure growth reduces $R$ because it enhances the signal. The distributions for other angular scales look qualitatively very similar.

\begin{figure}
  \centerline{\includegraphics[width=\hsize]{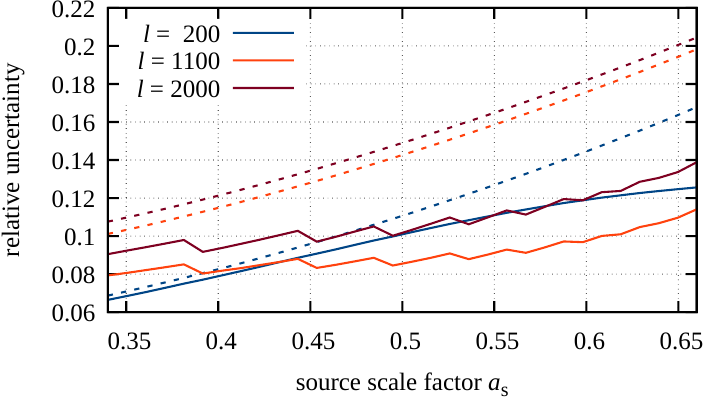}}
\caption{Integrated, one-dimensional, relative uncertainty $\bar R(a_\mathrm{s}, \ell)$ for the linearly and the non-linearly evolving power spectrum, as indicated, and for large ($\ell = 200$), intermediate ($\ell = 1100$) and small ($\ell = 2000$) angular scales.
{{The dashed lines correspond to linear structure growth whereas the solid lines take non-linear structure growth into account.}} 
For non-linearly growing cosmic structures, the curves vary little, increasing from {{$\approx9\,\%$ to $\approx 14\,\%$ for increasing source scale factor $a_s$}}. The uncertainty obtained with the linearly evolving power spectrum is higher than when non-linear evolution is assumed because non-linear structure growth substantially enhances the shear signal. The flatter slope of the curve for $\ell = 200$ and non-linear structure evolution is due to the transition of the power spectrum from linear to non-linear evolution illustrated in Fig.~\ref{fig:1}.}
\label{fig:6}
\end{figure}

The relative uncertainty integrated over perturbation scale factors $x$, i.e.\ the one-dimensional relative uncertainty $\bar R(a_\mathrm{s}, \ell)$, is shown in Fig.~\ref{fig:6}. Again, we show results there for the linearly and the non-linearly evolving power spectra, and for large and intermediate angular scales ($\ell = 200$ and $\ell = 1100$) in addition to small angular scales ($\ell = 2000$). As before, the source scale factor $a_\mathrm{s}$ varies between $1/3$ and $2/3$, corresponding to source redshifts between $z_s = 2$ and $z_s = 0.5$, respectively.

For linear structure growth and intermediate to small angular scales, the total relative uncertainty increases from {{$\approx10\,\%$ for sources at redshift $z_s \approx 2$ to $\approx20\,\%$}} for more nearby sources at redshift $z_s\approx0.5$. The main reason for this increase is that the cosmological weak-lensing signal is getting weaker as the source redshift decreases. With the non-linearly evolving power spectrum, the uncertainty is somewhat smaller, increasing from {{$\approx9\,\%$ for distant to $\approx14\,\%$}} for more nearby sources. This again reflects that the signal increases for non-linearly growing density fluctuations, reducing its relative uncertainty. The different shape of the curve for $\ell = 200$ and non-linear structure growth is due to the transition of the spectrum from linear to nonlinear behaviour, quantified by the function $\kappa(k, a)$ defined in (\ref{eq:16}) and shown in Fig.~\ref{fig:1}. The solid curves in Fig.~\ref{fig:6} show that the expansion function is empirically known well enough already to predict the cosmological weak-lensing power spectrum with a relative uncertainty of $\approx10\,\%$, without assuming any specific cosmological parameters apart from $\Omega_\mathrm{m0}$.

\section{Conclusions}
We have investigated how the power spectrum $C_\ell^\gamma$ of weak cosmological gravitational lensing changes with the expansion function $E(a)$ of the cosmic background. We are interested in this change for two main reasons: First, in view of a possible time dependence of dark energy, it may be important to know at which redshifts $C_\ell^\gamma$ is most sensitive to changes in the expansion function or, in other words, at which redshifts changes in the expansion function need to be for lensing to be most efficient in detecting them. Second, owing to a multitude of precise cosmological measurements, it has become possible to reconstruct the cosmic expansion function purely empirically, i.e.\ without reference to a specific cosmological model, with astonishing accuracy. Using this empirically determined expansion function for calculating the weak-lensing power spectrum, the remaining uncertainty of the expansion function will propagate into the power spectrum. In this context, it is interesting to see how accurately the weak-lensing power spectrum can be predicted based on a purely empirically determined expansion function.

Our main results are summarised in Figs.~\ref{fig:7} and \ref{fig:6}. The curves for non-linear structure growth in Fig.~\ref{fig:7} show that relative changes in the expansion function are amplified by a factor between {{$2$ and $6$}} in the cosmic-shear power spectrum, varying little with the scale factor $x$ where the expansion function is changed. The shear power spectrum is thus approximately equally sensitive to changes of the expansion function anywhere along the line-of-sight between the observer and the source, and it changes by {{$\approx 2-6\,\%$}} if the expansion function changes by $\approx 1\,\%$.

The corresponding curves in Fig.\ \ref{fig:6} show that the current uncertainty in the empirically determined expansion function plotted in Fig.\ \ref{fig:4} causes a relative uncertainty in the cosmic-shear power spectrum rising from {{$\approx9\,\%$}} for sources at redshift $z_\mathrm{s} \approx 2$ to {{$\approx 14\,\%$}} for sources at redshift $z_\mathrm{s} \approx 0.5$. In other words, if the cosmic expansion function underlying theoretical predictions of the weak-lensing power spectrum, which is usually taken from a cosmological model, would be substituted by its empirically determined counterpart, these theoretical predictions would be uncertain by about {{$10\,\%$}} for sources at redshift $z_\mathrm{s}\approx 1$. This may still seem quite large, but will shrink and improve as further cosmological distance measurements come in.

\section*{Acknowledgements}

This work was supported in part by Deutsche Forschungsgemeinschaft (DFG) under Germany's Excellence Strategy EXC-2181/1 - 390900948 (the Heidelberg STRUCTURES Excellence Cluster). {{We thank Alexander Oestreicher for corrections and helpful comments}.} 

\section*{Data Availability}

The data that support the plots within this paper and other findings of this study are available from the corresponding authors on reasonable request.



\bibliographystyle{mnras}
\bibliography{bibfile} 








\bsp	
\label{lastpage}
\end{document}